# Singular-phase nanooptics: towards label-free single molecule detection

V. G. Kravets[1], F. Schedin[1], R. Jalil[1], L. Britnell[1], R. V. Gorbachev[1], K. S. Novoselov[1], A. K. Geim[1], A. V. Kabashin[2] and A. N. Grigorenko[1]

[1]*School of Physics and Astronomy, University of Manchester, Manchester, M13 9PL, UK*

[2]*Laboratoire Lasers, Plasmas et Procédés Photoniques (LP3, UMR 7341 CNRS), Faculté des Sciences de Luminy, Aix-Marseille University, 163 Avenue de Luminy, 13288 Marseille Cedex 09, France*

**Non-trivial topology of phase is crucial for many important physics phenomena such as, for example, the Aharonov-Bohm effect [1] and the Berry phase [2]. Light phase allows one to create "twisted" photons [3, 4], vortex knots [5], dislocations [6] which has led to an emerging field of singular optics relying on abrupt phase changes [7]. Here we demonstrate the feasibility of singular visible-light nanooptics which exploits the benefits of both plasmonic field enhancement and non-trivial topology of light phase. We show that properly designed plasmonic nanomaterials exhibit topologically protected singular phase behaviour which can be employed to radically improve sensitivity of detectors based on plasmon resonances. By using reversible hydrogenation of graphene [8] and a streptavidin-biotin test [9], we demonstrate areal mass sensitivity at a level of femto-grams per mm$^2$ and detection of individual biomolecules, respectively. Our proof-of-concept results offer a way towards simple and scalable single-molecular label-free biosensing technologies.**

It is known that phase of light possesses nontrivial topology: it is a cyclic variable which is not defined at any point where the light intensity is zero (a point of darkness). Darkness in the real space can be obtained by using multiple beam interference [5], beams of higher transverse order [10] or near-fields [7]. For many applications, however, a singular behaviour of light phase in the spatial frequency domain (forming the basis of Fourier optics) is required. Prominent examples of such applications include high precision metrology and sensing, in which sharp phase features are employed to control stability of certain characteristics or the course of processes and reactions [11]. We demonstrate below that singular-phase behaviour can be achieved by using plasmonic nanostructures (see Fig. 1) and employ this feature to improve sensitivity of bio and chemical nanosensors based on optical transduction.

Optical transduction methods avoid expensive, time-consuming and precision-interfering labelling steps to mark analytes. Instead, they register the attachment of a ligand to its receptor via refractive index (RI) monitoring, which enables a real-time control of binding/recognition events [12, 13]. Surface plasmon resonance (SPR) forms the core of label-free optical transduction technology offering much superior sensitivity due to a strong electric field probing target molecules under conditions of resonant excitation of plasmons [14]. The spectacular progress of the SPR technology in recent years is strongly alimented by the development of numerous affinity models and protocols for gold surfaces. An extension of SPR called localized plasmon resonance (LPR) is realised by using metallic nanostructures [15], which makes possible a number of new functionalities including compatibility with modern bio-molecular nano-architectures [12], size selectivity and spectral tuneability [15, 16, 9], drastic field enhancement [17], nano-tweezing [18].

The advancement of plasmonic nanosensor technology towards detection of trace amounts of low molecular-weight analytes (drugs, toxins, etc.), such that only a few binding events are detected, is an appealing goal that can lead to a large impact in many fields of biomedicine, pharmacology and environmental safety [12, 13]. To achieve this goal, the plasmonic technology needs a drastic improvement in sensitivity. Although the existing LPR methods can detect 100-1000 molecules of relatively large analytes [15, 16], the detection limit in terms of the amount of a biomaterial accumulated on the surface is, typically, ~1000pg/mm$^2$, that is, much larger as compared to the conventional SPR (1pg/mm$^2$) [14] which in turn inferior by 3-4 orders of magnitude to labelling methods [12]. Ultimately, the low sensitivity problem of plasmonic transducers is connected to inherent losses in metallic nanostructures.

The plasmonic structures suggested in this work take advantage of enhanced phase sensitivity near phase singularities, which occur if light intensity sharply drops [19]. This behaviour has already been used to improve microscopy [20] and lower the detection limit of the SPR sensing technology by an order of magnitude [21, 11]. By designing nanomaterials with diffractive coupling of localised plasmons (DCLP), (theoretically suggested in [22, 23], first observed in [24] and independently confirmed in [25]) we solve the problem of inherent losses and create the complete darkness yielding to phase singularities. By using graphene hydrogenation, we estimate the detection limit of our nanomaterials at a level of ≈0.1fg/mm$^2$, which is 4 orders of magnitude better than reported in literature for SPR. We also show that suggested nanomaterials can be applied for biosensing and provide an unprecedented sensitivity in the absence of labels.

*How the nanomaterial works – coupling of localised plasmons.* Our devices consist of a regular array of submicron-scale structures made from Au (see Figs. 1a and 2). They display

LPR in the visible spectrum. If light interacts with such arrays in the reflection or attenuated total reflection geometry, this produces diffracted rays. The periodicity of our arrays is chosen in such a way that at a certain wavelength and a certain incident angle a diffracted ray becomes grazing and couples the LPR of individual nanostructures. This leads to a narrow collective plasmonic resonance [24] which is very sensitive to the environment [26] or binding events. Using diffractive coupled plasmons one can achieve an effective optical response which is normally not achievable in natural materials.

*Topologically protected darkness and phase sensitivity of coupled LPR.* Consider a light reflection from a thin film placed on a dielectric substrate. In the visible range, there exists a set of *n, k* (here $\hat{n} = n + ik$ is the refractive index of the film) for which the reflection is exactly zero. This set is shown by the solid brown curve in Fig. 1(c), where for concreteness the film thickness *d* is chosen to be 170nm, angle of incidence $\theta$=60° and the substrate is made of glass. In principle, it is possible to achieve these values of *n, k* by using a dielectric film near the Brewster angle. Although the enhanced phase sensitivity near the Brewster angle is used in Brewster angle microscopy [20] (and ellipsometry, in general), it is not widely used in biophysics since local electric fields for dielectric substrates are small. On the other hand, metal films can generate much stronger local fields due to plasmons and, therefore, provide a better phase sensitivity. Unfortunately, it is quite difficult to achieve phase singularity using a continuous metal film. For example, dispersion relations *n*($\lambda$), *k*($\lambda$) for gold yield the curve shown at the top of the image and result in non-zero reflection for gold films across the entire visible spectrum (measured ellipsometric reflection from a 170nm gold film is shown in the top panel of Fig. 1(c)).

The situation is different for a nanomaterial with DCLP. Using such plasmonic nanomaterials, one can manipulate effective $n_{eff}(\lambda)$, $k_{eff}(\lambda)$ and make them to intersect the zero reflection line in Fig. 1(c). The middle panel in this figure shows the effective dispersion curve and the measured reflection from the gold nanostripe structure schematically shown in Fig. 1(a) [27]. One can see a narrow plasmon resonance with the half-width of ≈12nm and quality of about $Q$~200. The detailed analysis shows that the light intensity reaches zero at certain wavelength and angle of incidence, which results in a singular behaviour of phase in the Fourier space. Indeed, the zero reflection line (the brown curve) separates two different regions in the ($n$, $k$) plane due to a nature of Fresnel reflection coefficients. Because the dispersion curve for the nanostructured gold starts in one of these regions and finishes in the other, it implies that it will always intersect the line of zero reflection curve due to the Jordan theorem [28] (which states that the line connecting two different regions separated by a boundary always intersects the boundary), see Fig 1(c). Relatively small imperfections or alterations in a structure will not change the fact that the dispersion curve for a nanostructured gold will connect two different regions in the ($n$, $k$) plane and hence the zero reflection for an altered structure will be still observed albeit at a slightly different wavelength. Therefore, the point of zero reflection for our nanomaterial is topologically protected due to the Jordan theorem. We will refer to this point as topological darkness.

Different excitation conditions require different structures to achieve topological darkness and phase singularity. Figure 2(a) shows a variety of unit cells for periodic nanostructures that we have experimentally employed to observe topological darkness in the Fourier space (zero reflection at certain angles and wavelengths). These include: nanodots (dot diameter about 100nm, thickness 90nm, array period about 300nm), double dots (which allow better control of the resonance position), gold dumbbells, stripes and arrays of holes in PMMA-gold double layers. The points of darkness allow one to achieve the increased phase sensitivity of affinity sensors based on coupled plasmons as discussed in refs. [11, 19, 21].

*Chemical sensing by using DCLP.* To evaluate the sensitivity of the suggested plasmonic structures to chemicals adsorbed at the surface, we employed hydrogenation of graphene [8], see Fig. 2(b). Graphene (with its well-defined 2D structure) was chosen as a test object because of a possibility to independently find the absorbed areal mass density of hydrogen. In addition, since graphene is easily functionalised we envisage that it may become a material of choice for calibration of plasmonic bio and chemical sensors. In our particular experiment, we used an array of double dots with overall sizes of 200x200 $\mu m^2$ (Fig. 2(c)). A graphene crystal of size 300x500 $\mu m^2$ was then transferred on top of the array (Fig. 2(d)), which was designed in such a way that the narrow diffractive coupled resonance and zero reflection occurred at 603nm at the angle of incidence ~69° (Fig. 3(a)). After the graphene transfer, we observed a red shift of the collective resonance to 612nm as shown in Fig. 3(a) due to the optical properties of graphene. Figure 3(b) plots changes in reflection due to the graphene hydrogenation in the vicinity of the collective resonance. The inset in Fig. 3(b) shows the evolution of the reflection minimum (related to changes in graphene's conductivity due to hydrogenation [8]) and the resonant wavelength (due to a change of the refractive index induced by adding hydrogen atoms). It is clear that the optical properties of DCLP are strongly affected by the graphene hydrogenation. Figure 3(c) shows the amplitude ratio for the *D-* and *G-* peaks in Raman spectra of graphene after its exposure to atomic hydrogen. We have used this ratio to evaluate percentage of absorbed hydrogen and, for example, it was ~1% after the first exposure (see ref. [8] and Methods).

Figure 3(d) shows our most important experimental result, a change of the ellipsometric parameters $\Psi$ and the phase $\Delta$ in the vicinity of the collective resonance after the hydrogenation exposure ($\tan(\psi)\exp(i\Delta) = r_p/r_s$, where $r_p$ and $r_s$ are reflection coefficients for *p*- and *s*-polarizations respectively). One can see that the phase changes by $\approx 44°$ (which is much larger than the associated relative change in $\Psi$). This change corresponds to a 1% hydrogen areal coverage, which translates to a mass density of <1pg/mm$^2$. The measured phase noise level for the experimental geometry was about 0.5° which gives the experimental areal mass sensitivity of <10fg/mm$^2$. If the optical system is thermally stabilized and advanced phase extraction methods are employed, a realistically achievable limit for phase noise could be as low as 0.005 degree [11]. In this case, the areal mass sensitivity could reach better that 100 atto-g/mm$^2$. It is also worth mentioning that the hydrogenation was reversible and all the reflection and Raman spectra returned to their original form after annealing (Fig. 3(b),(c)).

*Biosensing: streptavidin-biotin reaction observed by the singular-phase method.* To assess the applicability of our technique to biosensing, we used a well-developed and calibrated protocol based on the Streptavidin-Biotin affinity model [15] (Fig. 4). The surface of a nanodot plasmonic structure was functionalized by carboxylate groups and biotin was attached to carboxylate binding sites according to procedure described in Methods yielding to the attachment of up to 100 biotin molecules per each nanodot. Finally, the biotin-covered nanodots were exposed to 10pM Streptavidin (SA) solutions in 10mM phosphate-buffered saline for 3 hours, which resulted in their binding to all the biotin sites. As shown in Fig. 4(b)-(c), the attachment of SA led to changes in the phase of reflected light by ~25°. Note that this phase shift corresponds to the attachment of 20-100 SA molecules per nanodot (see Methods), which yields experimental sensitivity of 1-4 molecules per nanodot. As was demonstrated in [11], the resolution of phase measurements for thermally stabilized system with advanced phase detection can be better than $5·10^{-3}$ deg, which means that in principle one could resolve the attachment of 0.004-0.02 SA molecules per nanodot or <1 molecule attached per square micron area of our nanostructured devices. This detection limit is 2-3 orders of magnitude better than previously achieved for the conventional plasmonic nanosensors based on light intensity rather than phase changes [16].

To conclude, a careful design of plasmonic nanomaterials makes it possible to create topological darkness resulting in pronounced phase singularities. Such singularities are protected by topology from an external impact whereas diffraction-coupled plasmonic resonators allow extremely sharp plasmonic features. If employed for molecular recognition, the suggested plasmonic devices can provide unrivalled sensitivity on the single-molecule level, offering an alternative to the existing bio and chemical sensing technologies. The designed metamaterials allow their use within high throughput multi-sensing platforms and in combination with surface-enhanced techniques (fluorescence and Raman) techniques.

**Acknowledgements.**

We are grateful to the SAIT GRO Program, European Commission, and French National Research Agency (ANR).

**Methods**

1. **Graphene hydrogenation.** In order to bind hydrogen to the graphene surface we used a cold hydrogen dc plasma using a low-pressure (~0.1mbar) $H_2$/Ar (1:10) gas mixture. The plasma was ignited between Al electrodes ensuring the sample was at a safe distance (30 cm) from the discharge zone to avoid direct damage to the graphene lattice. We performed three plasma exposures of 20min each which provides a detectable level of single-sided hydrogenation [8]. The level of hydrogenation was estimated by measuring the *D* to *G* peak intensity ratio *I(D)/I(G)* in the Raman spectrum of the hydrogenated samples (we used a Renishaw RM1000 spectrometer with 514nm excitation wavelength). It has been demonstrated [29] that the ratio can be used to determine the typical distance between the defects $L_D$ using the following relation: $L_D = 4.24 \times 10^{-5} \lambda^2 \sqrt{I(G)/I(D)}$, where $\lambda$ is the wavelength measured in nanometers, *I(G)* and *I(D)* are the counts for *G* and *D* Raman peaks of hydrogenated graphene. This gives an estimate of $L_D \approx 50$nm after the first hydrogenation. However, taking into account the tendency for hydrogen atoms to form clusters on the surface of graphene, we have to assume that $L_D$ provides us only with the typical distance between the clusters of hydrogen atoms. To estimate the number of hydrogen atoms attached to graphene, we assume that the size of the cluster is smaller than the inter-cluster distance (5nm), which gives us an estimate of 1% hydrogenation. To check the reversible nature of hydrogenation we annealed the sample in nitrogen atmosphere for 4 hours at 200°C.

2. **Biosensing experiment.** We repeated the protocol described and calibrated in [15]. Glass slides with gold nanodots were incubated for 24 hours in 1 mM of 3:1 ethanolic solution of 1-octanethiol (1-OT) and 11 – Mercaptoundecanoic acid (11-MUA) purchased from Sigma-Aldrich yielding to the formation of a self-assembling monolayer with 10% surface coverage with carboxylate binding sites [15]. Since the active surface of each nanodot was equal to ~$10^4 nm^2$ such a procedure led to 2,000 active sites per nanodot. After incubation, the nanodots were rinsed with ethanol and dried in flowing nitrogen. Then, 1 mM biotin (Sigma-Aldrich) in 10 mM Phosphate-Buffered Saline (PBS) solution was linked to surface carboxyl groups using 1-ethyl-3-[3-dimethylaminopropyl] carbodiimide hydrochloride (EDC) coupling over 3 hours period. Taking into account ~1-5% efficiency of EDC coupling [15], up to 20-100 biotin molecules could attach to each nanodot. Finally, the biotin-covered nanodots were exposed to 10pM Streptavidin (SA, Sigma - Aldrich) solutions in 10mM PBS for 3 hours. Samples were finally rinsed with 10 mM PBS and water to remove all unspecifically bound molecules.

Excitation Energies. *Nano Letters* **11**, 3190-3196 (2011).

Figure Captions.

Fig. 1. **Singular phase and topologically protected darkness.** (a) Schematics of light reflection from a nanostructured Au film. (b) Calculated phase, $\Delta$, of the reflected light for the structure shown in (a) as a function of the wave-vector near the point of zero reflection. The blue line shows smooth phase behaviour far away from the singularity, while the yellow line shows the phase jump when a reflection curve goes through the point of rapid phase change. (c) The brown dispersion curve ($n(\lambda)$, $k(\lambda)$) corresponds to the line of zero reflection (phase singularity) for *p*-polarized light calculated from Fresnel coefficients for a 170nm film on a glass substrate, $\theta$=60°. The dispersion curve for standard bio-compatible plasmonic materials lies away from this singularity curve (for example, the top curve is for Au). The top panel shows the measured ellipsometric reflection $\Psi$ for a 170 nm Au film, which does not exhibit any darkness. The situation is different for nanostructured gold. In this case, the dispersion curve can pass through the brown curve of phase singularity and the reflection reaches exactly zero. The bottom curve plots the dispersion ($n(\lambda)$, $k(\lambda)$) for the nanostructure in (a) and the bottom inset shows the experimental behaviour of $\Psi$ exhibiting a topologically protected zero reflection at $\lambda$~630nm. The colour change along the plotted curves represents the corresponding colours of visible light.

Fig. 2. **Hydrogenation of graphene placed on top of a singular-phase nanostructure.** (a) Unit cells of various arrays that exhibited zero reflection and phase singularities in our experiments. (b) Schematically, a square array of Au double-dots on a glass substrate covered by a weakly hydrogenated graphene crystal. Localised

plasmon resonances of the Au dots are coupled by the grazing diffracted wave (the red arrow). (c) A scanning electron micrograph of such a nanostructure. (d) Its optical image.

Fig. 3. **Evaluation of sensitivity for singular-phase plasmonic detectors.** (a) Ellipsometric reflection spectra $\Psi$ in the region of the collective plasmon resonance for the pristine double-dot array (black curve) and with graphene transferred on top (red). The angle of incidence is 69°, the array constant $a$=320nm, the average size of the dots $d$=110nm, their separation $s$=140nm. The inset shows the entire spectrum for the pristine case. (b) Evolution of the $p$-polarized reflection of the structure in (a) during hydrogenation and annealing: the red curve corresponds to initial spectra; green - 20min of hydrogenation, blue - 60min, black - after annealing. Inset: changes in position and depth of the resonance. (c) The ratio of the amplitudes of D and G peaks in graphene as a function of hydrogenation. The inset shows a typical Raman spectra (40min of hydrogenation; the excitation wavelength $\lambda_{exc}$=514nm). (d) $\Psi$ and $\Delta$ for the cases of weakly hydrogenated (20min) and pristine graphene as a function of $\lambda$ (incident angle of 70°).

Fig. 4. **Biosensing with a plasmonic nanomaterial.** (a) Typical schematics of measurements. (b) $\Psi(\lambda)$ for the incidence angle of 53° and the array parameters $a$=320nm, $d$=135nm and $s$=140nm. (c) Evolution of $\Delta$ and $\Psi$ with time as SA molecules bind to functionalized Au dots ($\lambda$ =710 nm).

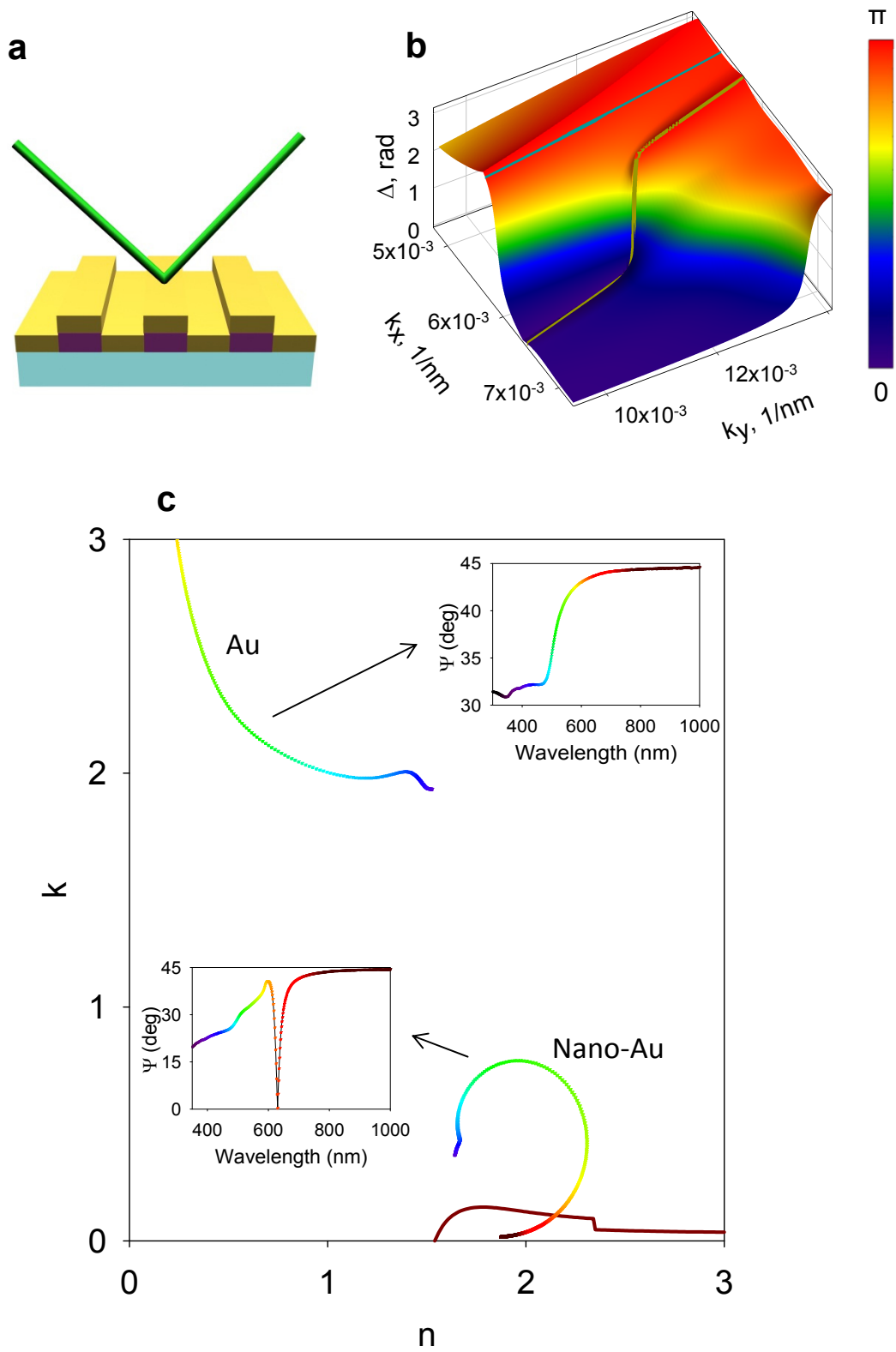

Fig. 1

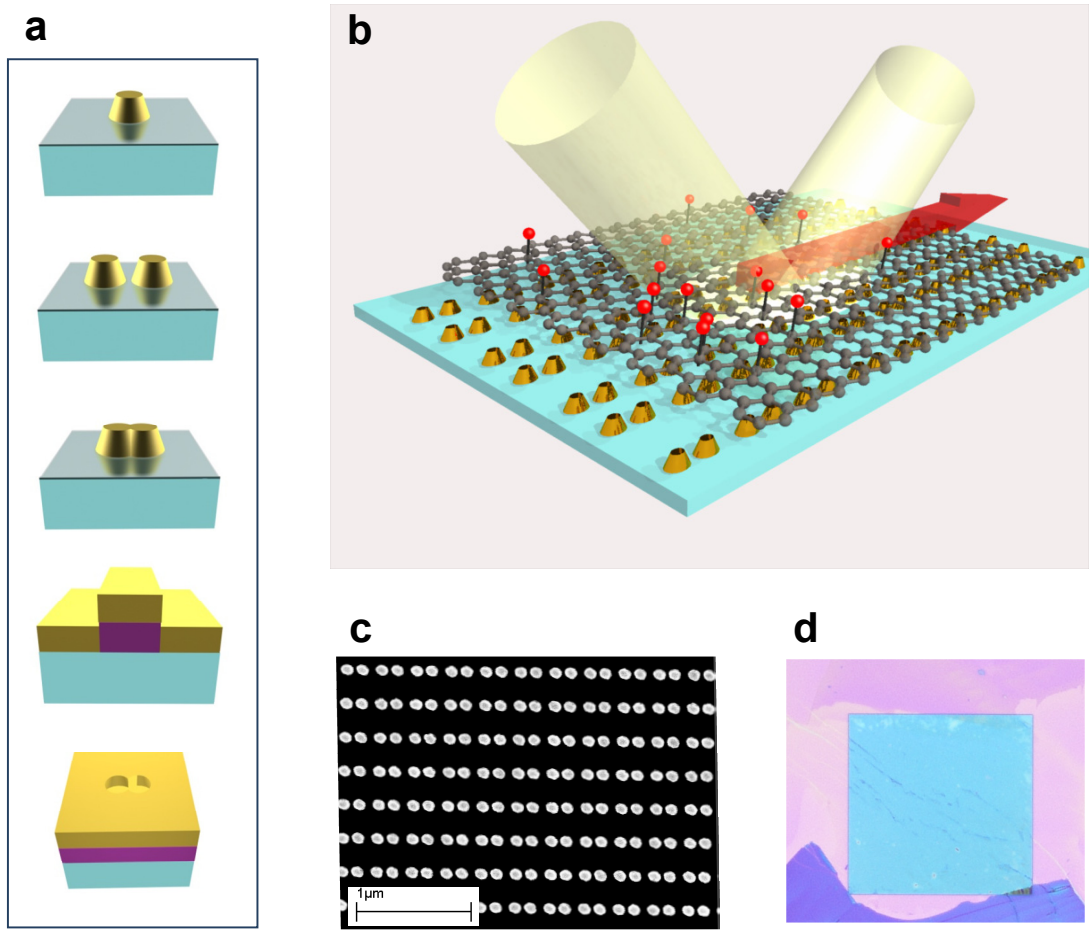

Fig. 2

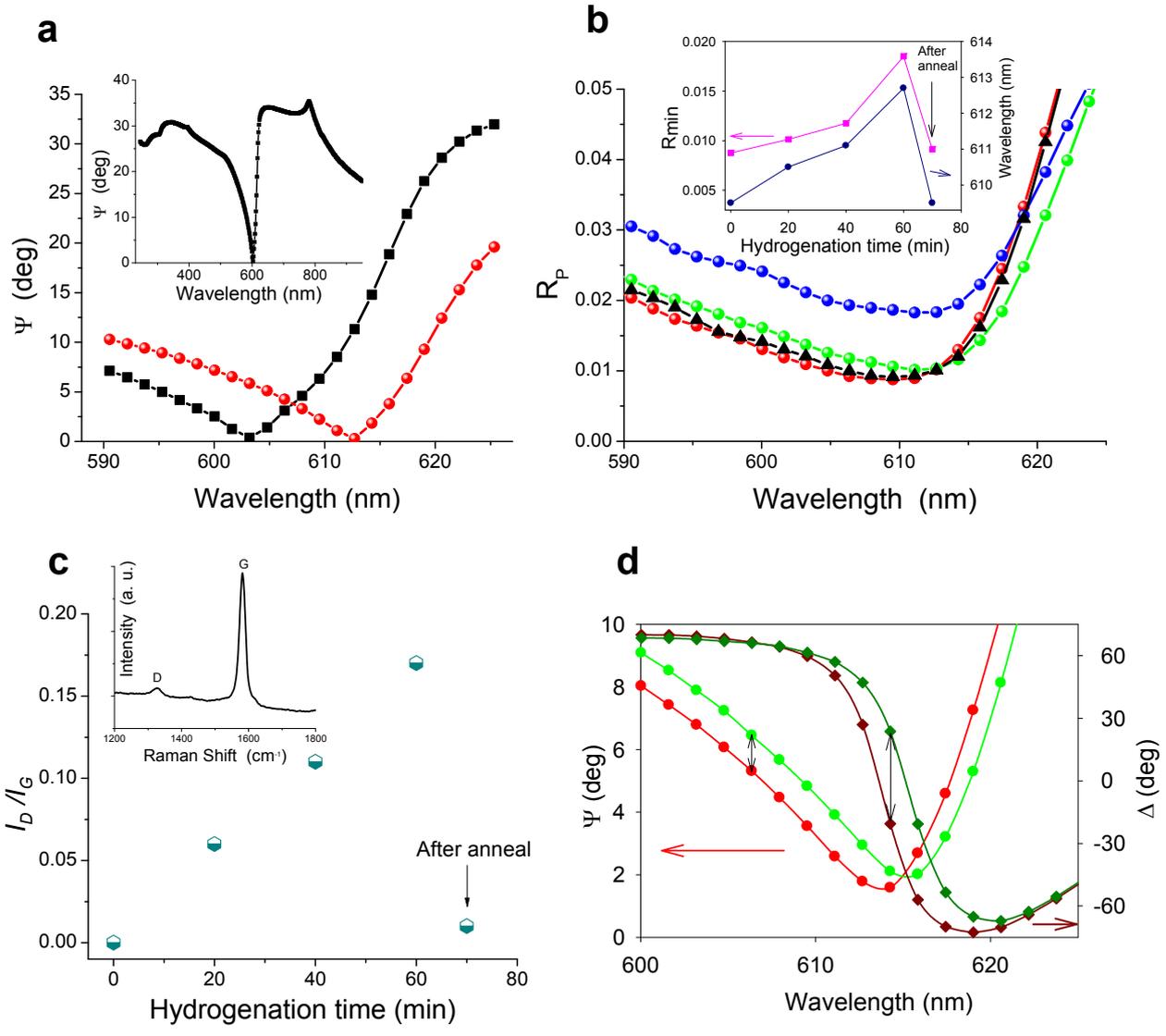

Fig. 3

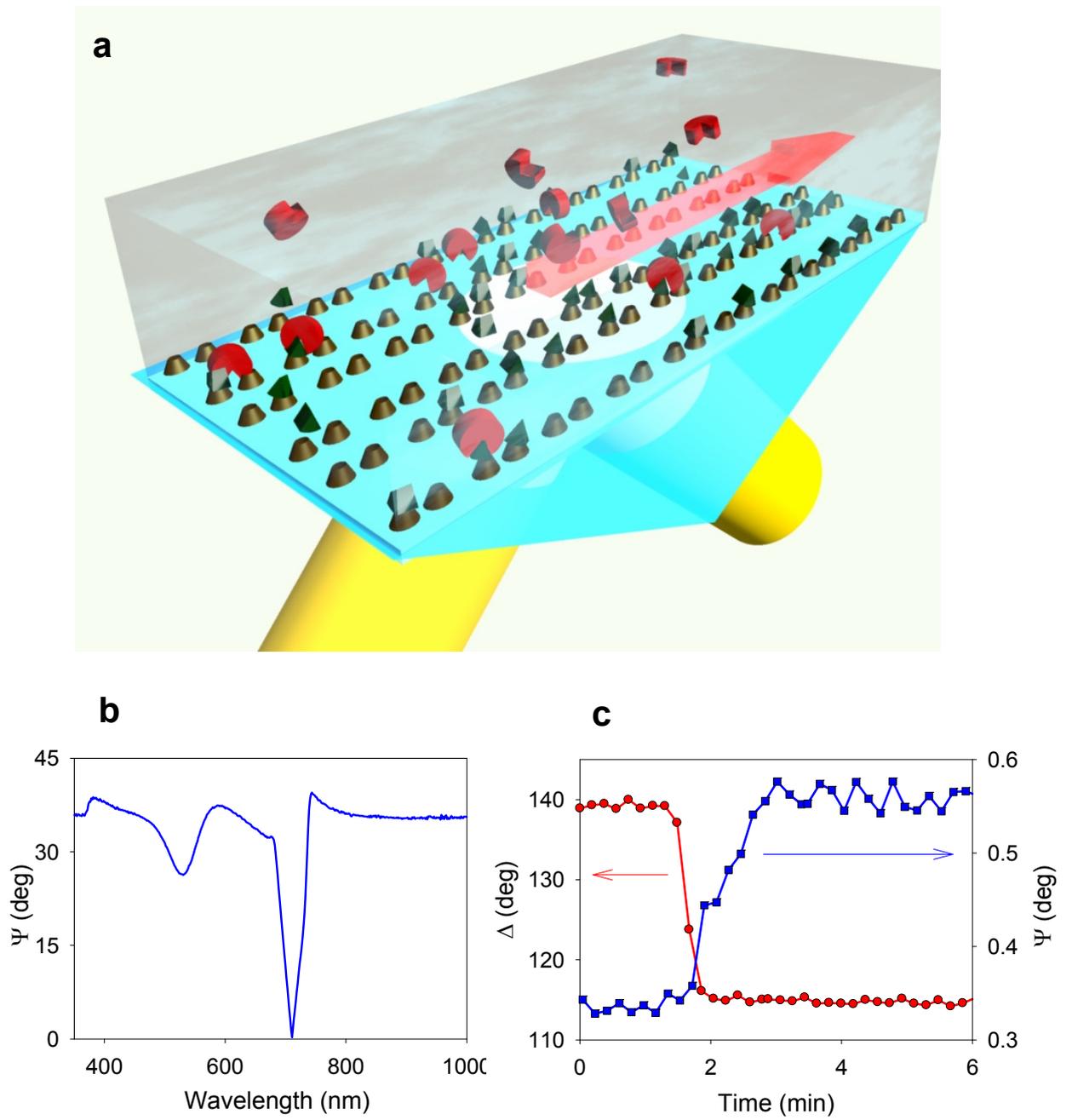

Fig. 4